\documentclass{jfm}

\usepackage{graphicx}

\newcommand{\Fr}{F\!r}

\newcommand{\ra}{\right >}
\newcommand{\la}{\left <}
\newcommand{\noi}{\noindent}
\renewcommand{\deg}{{}^{\circ}}

\newcommand{\dx}{\partial_x}
\newcommand{\dy}{\partial_y}
\newcommand{\dz}{\partial_z}
\newcommand{\dt}{\partial_t}
\newcommand{\dxx}{\partial_{xx}}

\title{Integral method for flows down an incline: viscous, turbulent and granular cases}
\author[A. Fourri\`ere, P. Claudin and B. Andreotti] {A\ls N\ls T\ls O\ls I\ls N\ls E \ns F\ls O\ls U\ls R\ls R\ls I\ls \`E\ls R\ls E ,  \ns P\ls H\ls I\ls L\ls I\ls P\ls P\ls E \ns C\ls L\ls A\ls U\ls D\ls I\ls N \ns \and B\ls R\ls U\ls N\ls O\ns A\ls N\ls D\ls R\ls E\ls O\ls T\ls T\ls I}
\affiliation{
Laboratoire de Physique et M\'ecanique des Milieux H\'et\'erog\`enes (PMMH),\\
UMR 7636 CNRS -- ESPCI -- Universit\'es Paris 6 et 7,\\
10 rue Vauquelin 75231 Paris Cedex 05, France.}

\pubyear{????}
\volume{???}
\pagerange{??--??}
\date{\today}
\begin{document}
\maketitle

\begin{abstract}
The integral method can be used to model accurately flows down an inclined plane. Such a method consists in projecting the full 3D equations on a lower dimensional representation. The vertical velocity profiles have their functional form fixed, based from the exact solution of homogeneous steady flows. This projection is achieved by integration of the momentum equation over the flow depth -- Saint-Venant approach. Here we generalize the viscous case to two non-newtonian constitutive relations: a Prandtl-like turbulent closure and a local granular rheology. We discuss one application in each case: the formation of anti-dunes in viscous streams, the transverse velocity profile in turbulent channels and the Kapitza instability in dense granular flows. They demonstrate the usefulness of this approach to get a model qualitatively correct, quantitatively reasonable and in which the dynamical mechanisms at work can be easily identified.
\end{abstract}

\section{Introduction}
The accurate hydrodynamical description of free surface flows, although initiated during the XIX$^{\rm th}$, still remains a modern problem of physics. The difficulties to be overcome are of different sorts. In some situations, like a viscous flow on a slope, the governing equations are both well posed and easy to integrate numerically. However, a numerical experiment, although giving access to any measurement, by contrast to actual experiments, does not provide in itself a physical understanding of the system. Most often, theories aiming to enlighten the physical mechanisms at work have been based on low dimensional model equations --~like the lubrication approximation for viscous flows (\cite{LL59}). In other situations, the numerical inegration of the governing equations is not accessible, due to the range of relevant length-scales and time-scales involved. For instance, the simulation of turbulent flows is currently limited to slightly less than four decades in space while the simulation of a river section would require at least $10^4$ times more mesh-points. Even for viscous flows, the presence of a moving contact line implies a divergence of the viscous stress regularised at the molecular scale. Five to six decades of length-scales being involved, the simulation of a simple millimetre scale moving droplet is beyond usual computer facilities. One trick is to introduce a sub-grid model, like for Large Eddy Simulations of turbulence. Another solution consists again in deriving a low dimensional model. In the turbulent case, one can use the shallow water equations without (\cite{LL59,W74}) or with a damping term (\cite{J02}). In the viscous case, the lubrication approximation can be extended to moving contact lines by introducing the slip length (\cite{SDFA06,SADF07}). Finally, there exist situations, like dense granular flows, for which the constitutive relations have not been derived from first principles and are still under debate (\cite{GDR04}). In the particular case of granular matter, the depth averaged dynamical equation --~hereafter referred as the Saint-Venant equation (\cite{SV43,SV50})~-- has been widely used to bypass the problem. Within this framework, instead of prescribing a stress/strain relation, one simply models the friction on the static bed over which grains are flowing (\cite{SH89,SH91,DAD99,P99b,PF02,MVBPBSY03}).

In any flow bounded by a free surface $h(x,y,t)$, the idea is to simplify the full set of equations governing the evolution of the velocity $u(x,y,z,t)$ (3D equations) into that for the depth averaged velocity $U(x,y,t)$ (referred to as (2+1)D equations). To perform this reduction under controlled approximations, one usually assumes that the flow is thin, i.e. that the longitudinal variations are negligible in comparison to transverse ones: ${\mathcal O}(\dx) \ll {\mathcal O}(\dz)$. This restricts the validity of the approach to small interface slopes, i.e. $| \dx h | \ll 1$. To go beyond  this restriction, methods in the spirit of the lubrication approximation have been developped to recover quantitative results for larger slopes. For instance, a long-wavelength expansion has been performed by \cite{S06} for wedge-like geometries, where the restriction is on the curvature of the interface, i.e. $| h \, \dxx h| \ll 1$. It yields to an equation that is well adapted to treat contact line problems (\cite{SADF07,DFSA07}).

However, by construction, the lubrication approximation (even extended to larger slopes) is lacking important dynamical mechanisms, namely inertial effects and longitudinal diffusion of momentum. It is therefore unable to give much insight into instability and pattern formation mechanisms. This is why a lot of numerical models, starting from a long-wave expansion, have been derived since the pioneering work of \cite{K48,KK49} in order to study development of waves at the surface of the fluid into large-amplitude strongly non-linear localized structures (\cite{PMP83}). Assuming that the height of the waves remains small in comparison to their wavelength, i.e. $| h- \la h \ra \! | \ll \lambda$, the interface modulations are expected to be governed by a Benney-like equation (\cite{B66}), where the temporal evolution of $h$ is expressed as a function of various algebraic powers of $h$ and its gradients. The most recent models (\cite{RM98,RM00,RM04}) accurately and economically predict linear and non-linear properties of film flows up to relatively high Reynolds numbers.

Such developments do not exist for real turbulent or granular flows. In this paper, we show how one can, in a general way, take into account all terms of Navier-Stokes equations, while keeping the Saint-Venant projection. This goal is achieved by looking for approximate solutions (called below `test functions') whose form is inferred from that of the exact steady and homogeneous solutions of the original equations, and with which we require that these equations are fulfilled on average only, i.e. integrated over the flow depth. Of course, we expect this approach to work well for situations which are not too far from the steady and homogeneous case.

We shall explain the precise development of this integral method in the next section for the case of the well studied viscous flow. We then turn to more original calculations with the treatment of turbulent and granular flows. For each of these three cases, we illustrate our approach with a concrete example of application: the formation of dunes in viscous streams, the transverse velocity profile in turbulent channels and the Kapitza instability in dense granular flows. Although we are greatly interested in the physical insights of these different situations, we shall here focus on the technical discussion of the benefits of this approach.

\section{Viscous flow equations}

\begin{figure}
\centering
\includegraphics{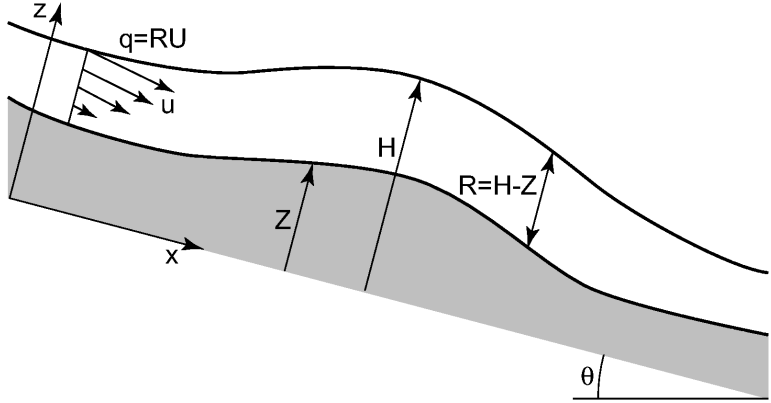}
\caption{Schematic drawing of the flow down an inclined of slope $\tan\theta$. $x$ is along the bottom, $z$ perpendicular to it. $y$ is in the third direction. $Z(x)$ is the equation of the bed and $H(x)$ that of the free surface. We note $R=H-Z$ the local fluid depth. $q$ is the local flow rate -- see equation (\ref{defq}) -- so that $U=q/R$ is the mean velocity at position $x$.}
\label{SchemaVisc}
\end{figure}

\subsection{Governing equations}
We consider the generic situation of a viscous fluid flowing down a plane inclined at an angle $\theta$ with respect to the horizontal. The fluid motion is referred to a Cartesian coordinate system in which the $x$-axis coincides with the inclined bottom. The $z$-axis is perpendicular to this plane and $y$ is in the plane, transverse to the slope. A schematic drawing of the flow with all notations is shown in figure \ref{SchemaVisc}. Combined with the incompressibility equation, the Navier-Stokes equation provides the theoretical framework that describes the motion of the fluid:
\begin{eqnarray}
\partial_j u_j & = & 0, \nonumber \\
\dt u_i + (u_j \partial_j) u_i & = & - \partial_i p + g_i + \nu \Delta u_i,
\label{NS}
\end{eqnarray}
where $\vec{u}= (u,v,w)$ is the velocity field, $p$ is the pressure divided by the fluid density $\rho$, $\vec{g}$ the gravity vector and $\nu$ the kinematic viscosity. Note that in the entire paper, we assume the density to be constant and omit this $1/\rho$ factor for simplicity. However, we shall still use the word `pressure' for $p$ -- same for momentum and stress. The boundary conditions we wish to impose are the following:\\
(i) all velocity components vanish at the bottom: $u_i(z=Z)=0$,\\
(ii) the free surface follows the fluid motion,
\begin{equation}
\dt H + u \, \dx H + v \, \dy H - w = 0,
\end{equation}
all velocity components $u$, $v$ and $w$ being evaluated at $z=H$,\\
(iii) the stress normal to the surface vanishes.

\subsection{Exact homogeneous steady solutions}
The strategy of the integral method proposed here is to solve exactly the case of homogeneous steady solutions and to use it as a test function to derive the Saint-Venant equations. In the viscous case, equation (\ref{NS}) admits a trivial solution corresponding to a steady constant-thickness film. Assuming that all derivatives with respect to $t$, $x$ and $y$ vanish, there is only a streamwise velocity $u$ and the equations reduce to:
\begin{eqnarray}
0 & = & \nu \, \partial_{zz} u + g \sin \theta, \nonumber \\
0 & = & -\partial_z p -g \cos \theta,
\end{eqnarray}
which, for a film of thickness $H_0$ (in this homogeneous case, $Z_0=0$ and thus $R_0=H_0$), yields to
\begin{eqnarray}
u_0(z) & = & \frac{g \sin \theta}{2\nu} \, z (2 H_0-z), \nonumber \\
p_0(z) & = & p_{\rm atm} + g (H_0-z)\cos \theta,
\end{eqnarray}
where $p_{\rm atm}$ is the reference pressure.

\subsection{Integral method}
The next step is to reduce the full equations to a set of two equations governing the evolution of the local depth of the flow $R = H - Z$. For the sake of simplicity we shall keep to $1+1$ dimensions, but the following derivation can be easily generalized to (2+1)D. The local flow rate $q$ is defined as:
\begin{equation}
q=\int_Z^H u \, dz.
\label{defq}
\end{equation}
From $q$, we define the mean velocity $U=q/R$. The first of these equations is simply the (integral) mass conservation: $\dt R + \dx q = 0$. The second one is the projection of the Navier-Stokes equation integrated over the height of the flow. It gives an integral momentum conservation equation (Saint-Venant equation):
\begin{equation}
\int_Z^H \left [ \dt u + u \dx u + w \dz u \right ] \, dz =
\int_Z^H \left [ g \sin\theta - \dx p + \nu \Delta u \right ] \, dz.
\label{StVenant}
\end{equation}
In order to compute these integrals, the vertical profiles are assumed to remain close to those obtained in the homogeneous steady case. We thus take a parabolic velocity test profile parametrized by $q$, $R$ and $Z$ (or $H$)
\begin{equation}
u = 3 \, \frac{q}{R} \left( \frac{z-Z}{R} \right) \left (1-\frac{z-Z}{2R}\right ).
\label{uparabol}
\end{equation}
Similarly, the pressure profile remains hydrostatic:
\begin{equation}
p = p_{\rm atm} + g (H-z)\cos \theta.
\end{equation}
We shall turn to the limit of this assumption in the conclusion.

Because of the incompressibility relation, we can compute the $z$-component of the velocity:
\begin{equation}
w = -\int_Z^z \dx u \, dz.
\end{equation}
An analysis of the order of magnitude of this velocity leads, as expected, to a ratio $w/u \sim \dx Z$.  In the following we shall then focus on the streamwise component velocity only, as we are interested in situations with a weak bed slope correction to the homogeneous $Z(x)=0$ case. Using incompressibility as well as the free surface and bottom boundary conditions, one can write the inertial terms as the derivative of the momentum flux:
\begin{equation}
\int_Z^H \left [ \dt u + u \dx u + w \dz u \right ] \, dz =
\dt \int_Z^H u \, dz + \dx \int_Z^H u^2 \, dz=
\dt q + \dx \left ( \alpha \frac{q^2}{R} \right ) .
\end{equation}
Interestingly, the value of $\alpha$ depends only on the test function and is thus determined in the homogeneous steady case: $\alpha=6/5$ for the parabolic profile (\ref{uparabol}).

The right hand side of the Saint-Venant equation contains the different forces integrated vertically, namely:\\
(i) the projection of gravity along the streamwise direction, $gR\sin\theta$,\\ (ii) the pressure gradient, $-gR \cos\theta \dx H$,\\
(iii) the vertical diffusion of momentum,
\begin{equation}
\int_Z^H \partial_z (\nu \partial_z u) \, dz=-\nu \partial_z u \arrowvert_{z=Z} = -3 \nu \frac{q}{R^2},
\nonumber
\end{equation}
(iv) the streamwise diffusion of momentum,
\begin{equation}
\int_Z^H \dx (\nu \dx u) \, dz=\nu \dxx q - 3 \nu \left[ \frac{q}{2R} \, \dxx H + \dx \left (\frac{q}{R}\right )\dx H + \frac{q}{R^2}\left (\dx Z\right )^2\right].
\label{streamwisediffmomentum}
\end{equation}
We finally obtain the following system of equations:
\begin{eqnarray}
Z+R&=&H,\\
\dt R + \dx q & = & 0 \nonumber \\
\dt q + \dx \left (\alpha \frac{q^2}{R}\right ) & = & gR\cos \theta\;(\tan \theta - \dx H)- 3 \nu \frac{q}{R^2} \left(1+ \frac{1}{2} R \dxx H\right)+\nu \dxx q , \nonumber \\ 
&-& 3 \nu \left[ \dx \left (\frac{q}{R}\right )\dx H + \frac{q}{R^2}\left (\dx Z\right )^2\right].
\label{viscousSV}
\end{eqnarray}
In the long wavelength approximation, at low Reynolds number, one can neglect, in equation (\ref{viscousSV}), its left-hand side as well as the terms coming from (\ref{streamwisediffmomentum}). One then recovers the standard lubrication equations. In the right-hand side of equation (\ref{streamwisediffmomentum}), one recognizes a term corresponding to the viscous diffusion of $q$. The three other dissipative terms originate from the geometry of the flow and their interpretation is less straightforward.

The above equation set is very close in spirit to the derivation of \cite{S67}. Other versions of this modeling have been proposed for specific problems (\cite{RM98,RM00,RM04}). They all keep the same functional form, but with different numerical coefficients, adjusted to match experimental data. By contrast, our equations are less quantitative but directly come from the original Navier-Stokes equations and the different terms have therefore not been modified in an \textit{ad hoc} way for some particular purpose. All dynamical mechanisms are thus present and can be backtracked to get their physical meaning. The interest of the terms extending the lubrication approximation for the problem of Kapitza instability and roll waves has been discussed at length by \cite{RM98}. We refer the interested reader to the review section of that article. In the following we discuss another example: the formation of sand anti-dunes in a viscous stream.

\subsection{Steady flow over a bumpy bottom}
Let us first apply our approach to the simple case of a steady flow over a non-uniform bed $Z(x)$. From conservation of mass, we infer that the flow rate is uniform along the stream and is equal to:
\begin{equation}
q_0= H_0 U_0 = \frac{1}{3} \, \frac{g  H_0^3 \sin \theta}{\nu}
\end{equation}
The flow may be usefully characterized by two independent non-dimensional numbers among the following three: the slope $\tan \theta$, the Froude number
\begin{equation}
\Fr^2 = \alpha \, \frac{U_0^2}{gH_0\cos\theta}
\end{equation}
and the Reynolds number
$$Re = \frac{U_0 H_0}{\nu} = \frac{5}{2} \, \frac{\Fr^2}{\tan\theta}$$

In the limit of a small bed amplitude, one can without loss of generality consider sinusoidal perturbations of the form $e^{i\bar{k}x/H_0}$, where $\bar{k}=k H_0$ is the wavenumber rescaled by the flow height. We note $q_1$, $Z_1$, $R_1$ and $H_1$ the respective disturbances of $q$, $Z$, $R$ and $H$. At the linear order, we get:
\begin{eqnarray}
Z_1+R_1 & = & H_1 \nonumber\\
q_1 & = & 0 \nonumber\\
-i\bar{k} \Fr^2 R_1 & = & \tan\theta R_1 -i\bar{k} H_1 +2 \tan\theta R_1 +\frac{1}{2} \tan\theta \bar{k}^2 H_1
\end{eqnarray}
Combining the first and last equations, we obtain the deformation of the free surface as a function of that of the bed.
\begin{equation}
H_1 = \frac{3 \tan\theta + i\bar{k} \Fr^2}{(3 + \frac{1}{2}\bar{k}^2)\tan\theta +i\bar{k} (\Fr^2-1)} \, Z_1.
\end{equation}
%

\begin{figure}
\centering
\includegraphics{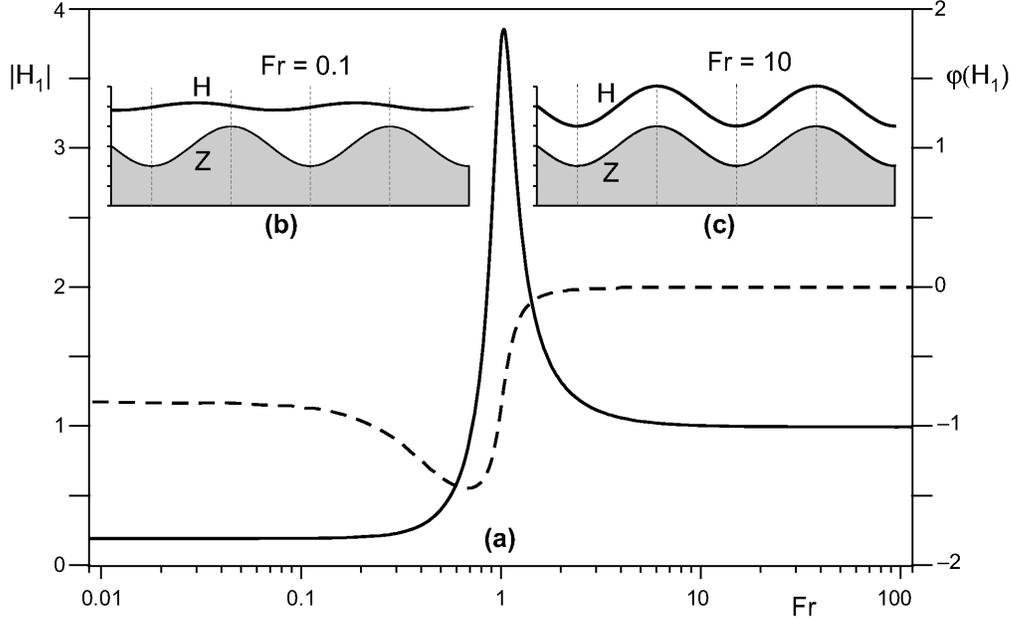}
\caption{(a) Amplitude (solid line) and phase (dashed line) of the free surface with respect to the bed deformations as a function of the Froude number, for $\tan\theta = 0.1$ and $\bar{k} = \pi/2$. The free surface profiles have been plotted for a low Froude number (b) and for a high one (c).}
\label{PhaseEtAmplitudeVisc}
\end{figure}

In figure \ref{PhaseEtAmplitudeVisc}, we plot the amplitude and the phase of the free surface profile with respect to the sinusoidal bed deformations as functions of the Froude number. Due to viscous damping, the bed and the free surface are out of phase essentially close to a unit Froude number, and the amplitude of the perturbation is strongly reduced. In contrary, for large Froude numbers, the disturbances of the free surface profile exactly match that of the bed profile in phase as well as in amplitude.

\begin{figure}
\centering
\includegraphics{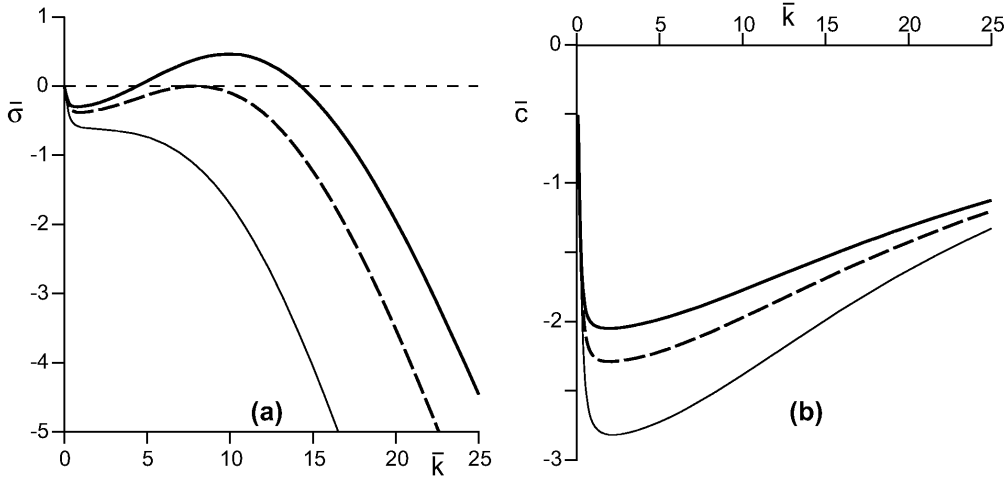}
\caption{Temporal growth rate $\bar{\sigma}$ (a) and phase velocity $\bar{c}$ (b) as functions of the wavenumber $\bar{k}$ for $\tan\theta = 0.05$ and $L_{\rm sat}/H_0 = 0.06$ at three different Froude numbers: $\Fr = 1.3$, $\Fr = \Fr_c$ and $\Fr = 1.4$ (thin solid, bold dashed and bold solid lines respectively).}
\label{SigmaKVisc}
\end{figure}

\subsection{Application: anti-dunes in a viscous stream}
We can use the previous calculation to predict the formation of sand dunes in a viscous stream. The following calculation is only an illustration of the integral method. We thus restrict ourselves to exhibit the interest of the new equations in comparison to the lubrication approximation. We refer the reader interested in dunes generated by a steady viscous flow to \cite{Y77,Y85,KLMC95,CM96,R97,CE00,VL05,KL05,CH06,C06}. The linear stability analysis of a flat sand bed can be described in the framework developed by \cite{ACD02,ECA05,CA06} for dunes forming in a turbulent boundary layer. The evolution of the bed $Z$ is governed by the sand mass conservation:
\begin{equation}
\dt Z + \dx Q = 0,
\label{exner}
\end{equation}
where $Q$ is the volumic sediment flux, i.e. the volume of the sand grains per unit time which cross an infinite normal surface of unit transverse width. The sand flux $Q$ itself depends on the basal shear stress $\tau$, with a space lag $L_{\rm sat}$ (\cite{ACD02})
\begin{equation}
L_{\rm sat} \, \dx Q = Q_{\rm sat} - Q,
\label{sat}
\end{equation}
where the saturated sand flux $Q_{\rm sat}$ is a function of $\tau$ only. For the following analysis, we do not need to know the details of the transport law $Q_{\rm sat}(\tau)$ but only the value of this saturated flux over a flat bed, $Q_0$, and that of the coefficient:
\begin{equation}
\beta=\frac{1}{Q_0} \frac{dQ_{\rm sat}}{d\tau} \, .
\end{equation}
The unstable modes are now of the form $e^{(\sigma-i\omega) t +i\bar{k}x/H_0}$. $\sigma$ is the temporal growth rate of the dunes and $c=\omega/k$ is their phase velocity. Equation (\ref{exner}) then reads $(\sigma-i\omega) Z_1+i\bar{k}/H_0 Q_1=0$. Similarly, equation (\ref{sat}) simplifies into $i\bar{k}/H_0 L_{\rm sat} Q_1= \beta Q_0 \tau_1 - Q_1$, where subscripts $1$ denote first order corrections. Combining these two equations, we get
\begin{equation}
\sigma-i\omega = - \frac{i\bar{k} Q_{0} \beta}{(H_0+i\bar{k}L_{\rm sat})} \frac{\tau_1}{Z_1}
\label{sigma_visc}
\end{equation}
which relates $\sigma$ to the ratio $\tau_1/Z_1$, governed by the hydrodynamics. We can see as from now that the sand bed instability can be related to the phase of the shear stress with respect to the bed (\cite{ECA05,CA06}).

To determine $\tau_1/Z_1$, we assume that the time scales over which the bed and the flow evolve are well separated. For the hydrodynamics, the sand bed can thus be considered as static, as in the previous section. One can compute the basal shear stress from equation (\ref{uparabol}) with $\tau = \nu \left . \dz u \right |_{z=0} = 3\nu q/R^2$. Its first order perturbation $\tau_1$ then reads:
\begin{equation}
\tau_1 = - 2 g \sin \theta R_1 =g \sin \theta \frac{\tan \theta \bar{k}^2-2 i \bar{k}}{(3+\frac{1}{2}\bar{k}^2)\tan\theta +i\bar{k} (\Fr^2-1)} \, Z_1 .
\end{equation}
Inserting this hydrodynamical relation into the expression of the growth rate (\ref{sigma_visc}),  we finally obtain:
\begin{eqnarray}
\frac{H_0}{Q_{0} \beta g \sin\theta} (\sigma-i\omega)
& \equiv & \bar{\sigma} - i\bar{\omega} \nonumber \\
& = & - \frac{(2+i \tan \theta \bar{k}) \bar{k}^2}{(1+i\bar{k}L_{\rm sat}/H_0) \left[(3+\frac{1}{2}\bar{k}^2)\tan\theta+i\bar{k} (\Fr^2-1)\right]} \, ,
\end{eqnarray}
where $\bar{\sigma}$ and $\bar{\omega}$ are the dimensionless growth rate and frequency as defined above. We also write $\bar{c}=\bar{\omega}/\bar{k}$.

\begin{figure}
\centering
\includegraphics{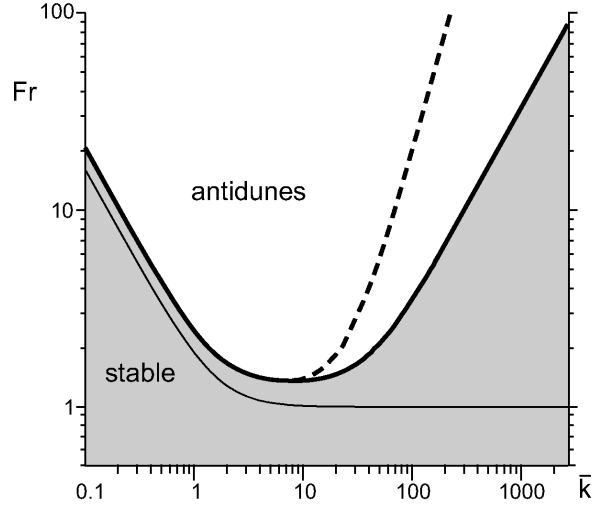}
\caption{Diagram of stability for such instability, where $\Fr$ is plotted as a function of $\bar{k}$ for the following parameter values: $\tan\theta = 0.05$ and $L_{\rm sat}/H_0 = 0.06$. Above the line, the growth rate is positive and the perturbation is unstable. The dashed line represents the most unstable mode. The thin line corresponds to the very same equations but without the terms coming from (\ref{streamwisediffmomentum}). In this case, there is no upper cut-off for $\bar{k}$ and the most unstable mode is at $\bar{k} \to \infty$.}
\label{FrVsKvisc}
\end{figure}

Both $\bar{\sigma}$ and $\bar{c}$ are represented as a function of $\bar{k}$ in figure \ref{SigmaKVisc} for different values of the Froude number. For $\Fr \le \Fr_c$, the perturbation remains stable for all wavelengths ($\bar{\sigma}$ is negative). Above $\Fr_c$, the perturbation is unstable for a range of $\bar{k}$. These lower and upper cutoffs are represented in the stability diagram of figure \ref{FrVsKvisc}. Note that the value of $\Fr_c$ depends on the other parameters $\tan\theta$ and $L_{\rm sat}/H_0$. Unlike the growth rate, the velocity of the waves is always negative (opposite to the fluid one) below or above the critical Froude number. As these sand dunes propagate up-stream, they correspond in fact to `antidunes'. Performing the very same calculation without the terms coming from (\ref{streamwisediffmomentum}), one get a similar scenario as that displayed in figure \ref{SigmaKVisc} with the important difference that when perturbations are unstable (above $\Fr_c=1$ independently of the other parameters), the corresponding range of wavenumbers extends up to infinitely large $\bar{k}$ (no upper cut-off). As a conclusion, the longitudinal diffusion of momentum plays a crucial role in the control of this instability. Our equations are thus able to recover the correct dynamical mechanisms that were present in the original 3D equations but were lost in the shallow water approximation --~and {\it a fortiori} in the lubrication approximation.

\section{Turbulent flow: the Prandtl mixing length approach}

In the previous section, we have shown how it is possible to extend Saint-Venant equations beyond the classical shallow water approximation. The method was illustrated on the extensively studied viscous case, which extends the lubrication equations to slightly inertial flows. We now apply the same method to fully turbulent flows, for which equation (\ref{StVenant}) strongly under-estimates the rate of energy dissipation.

\subsection{Reynolds stress and Prandtl mixing length closure}
At large Reynolds numbers fluctuations can no longer be neglected. Writing the Navier-Stokes equation for the mean flow (intended as an ensemble average over realizations), a turbulent stress $\tau_{ij}=- \la u'_i u'_j \ra$ adds to the viscous one, called the Reynolds stress tensor (\cite{LL59}). For large values of $Re=UH/\nu$, the viscosity term may be neglected and the Saint-Venant equation becomes:
\begin{equation}
\int_Z^H
[ \dt u + u \dx u + w \dz u ] \, dz
=
\int_Z^H
[ g \sin\theta - \dx p + \dx \tau_{xx} + \dz \tau_{xz} ] \, dz.
\end{equation}
 In the case of a homogeneous turbulent boundary layer, $\tau_{ij}$ can be related to the velocity gradient using dimensional analysis. The only length-scale is given by the distance to the boundary $z$ and the only time scale by the inverse of the shear rate $|\dot{\gamma}|$. The shear stress is then of the form $\ell^2|\dot{\gamma}| \dot{\gamma}$, where  $\ell = \kappa (z+z_0)$ is the Prandtl mixing length. $\kappa\simeq0.4$ is the von K\'arm\'an constant, determined experimentally, and $z_0$ is the hydrodynamic roughness. For a homogeneous shear stress $\tau_{xz}$, one obtains a logarithmic velocity profile 
 \begin{equation}
 u = \frac{\sqrt{\tau_{xz}}}{\kappa} \ln \left ( 1 + \frac{z}{z_0} \right ).
 \end{equation}
 
In the general case, there exists no univoque relation between the Reynolds stress $\tau_{ij}$ and the velocity field. However, situations close to the homogeneous turbulent boundary layer can be reasonably described by a phenomenological closure. For the sake of simplicity, we have chosen here to introduce a turbulent viscosity $\nu_t$, derived from the Prandtl mixing length approach. The Reynolds stress tensor is written as a function of the velocity gradient: $\tau_{ij} = \ell^2 | \partial_j u_i | \partial_j u_i $.  In the case of a river, it is commonly assumed that the pressure field is hydrostatic and that the velocity profile, as in the case of the turbulent boundary layer, is logarithmic. We thus write
\begin{eqnarray}
\label{uTurbul}
u & = & \frac{\sqrt{C_z}}{\kappa} \, \frac{q}{R} \, \ln \left ( 1 + \frac{z-Z}{z_0} \right ),\\
\label{pTurbul}
p & = & p_{\rm atm} + g (H-z) \cos \theta,
\end{eqnarray}
where $q$ is the instantaneous local flow rate defined in (\ref{defq}). The vertical integration of this profile $u(z)$ gives for self-consistency
\begin{equation}
\sqrt{C_z} = \frac{\kappa R}{(R+z_0)\ln \left ( 1 + \frac{R}{z_0} \right ) -R} \, .
\label{Chezy}
\end{equation}
This so-called Chezy coefficient is an important quantity as it gives the ratio of the shear stress on the bed to the square of the mean velocity: $\tau_{xz}(z=0) = C_z (q/R)^2$ (see below).

In the steady case, due to the linearity of the shear stress with $z$, we get the following expression for the Prandtl mixing length:
\begin{equation}
\ell = \kappa (z-Z+z_0) \sqrt{\frac{H-z}{R}} \, .
\end{equation}
We recover, with this phenomenological expression, that the size of the mixing eddies is limited by the distance to the rigid boundary as well as by that to the free surface. By analogy with the previous viscous equations, we introduce a turbulent viscosity $\nu_t$ in the definition of the shear stress tensor $\tau_{ij} = \nu_t \partial_j u_i$. In contrast to the viscous case where the viscosity is a constant, $\nu_t$ depends on $x$, $z$ and $t$:
\begin{equation}
\nu_t = \ell^2  | \partial_z u | = \kappa \sqrt{C_z} \, q \, \frac{(z-Z+z_0)(H-z)}{R^2} \, .
\end{equation}
%

\subsection{Integral method}
We consider a 2D turbulent fluid flow down an incline. Keeping the relations (\ref{uTurbul}) and (\ref{pTurbul}), we can express the components of the stress tensor as follows:
\begin{eqnarray}
\tau_{xx} = & -\tau_{zz} = & \left \{
\frac{1}{2}
\dx \left [ C_z \left (\frac{q}{R}\right )^2 \right ]
(z-Z+z_0) \, \ln \left (1+\frac{z-Z}{z_0}\right )
- C_z \left (\frac{q}{R}\right )^2 \dx Z
\right \} \frac{H-z}{R} \, , \nonumber \\
\tau_{xz} = & \tau_{zx} = &
C_z \left (\frac{q}{R}\right )^2 \frac{H-z}{R} \, .
\end{eqnarray}
Summing up all contributions, the turbulent Saint-Venant equation (\ref{viscousSV}) reads
\begin{eqnarray}
R + Z & = & H, \nonumber \\
\dt R + \dx q & = & 0, \nonumber \\
\dt q + \dx \left ( \alpha \frac{q^2}{R} \right )
& = &
g R \cos \theta \, (\tan\theta - \dx H)
- C_z \frac{q^2}{R^2}
- \frac{1}{36} \left [ R^2 (5/2 - 3\lambda)
\, \dxx \left ( C_z \frac{q^2}{R^2} \right )
\right . \nonumber \\
& + &
18 C_z \frac{q^2}{R^2} [R \dxx Z + 2(\dx Z)^2 + \dx R \, \dx Z] \nonumber \\
& + &
\left .
2 R \, [9\dx Z + (1-3 \lambda) \dx R]
\, \dx \left ( C_z \frac{q^2}{R^2} \right ) \right ],
\end{eqnarray}
with $\lambda = \ln \left ( 1 + \frac{R}{z_0} \right )$.

In the long wavelength approximation, one can neglect in the above equation the terms corresponding to longitudinal `diffusion' of momentum. Furthermore, under the approximation $z_0 \ll R$, the value of $\alpha \simeq 1 + 1/(\lambda-1)^2$ is close to unity. Therefore, one recovers the standard shallow water equation:
\begin{equation}
\dt q + \dx \left ( \frac{q^2}{R} \right ) = g R \cos \theta \, (\tan\theta - \dx H) - C_z \frac{q^2}{R^2} \, .
\label{shallowwater}
\end{equation}
Among the additional terms, one recognizes a term corresponding to a diffusion of momentum along $x$. As in the viscous case, the other dissipative terms originate from the geometry of the flow.

\subsection{Example: transverse velocity profile in a river}

\begin{figure}
\centering
\includegraphics{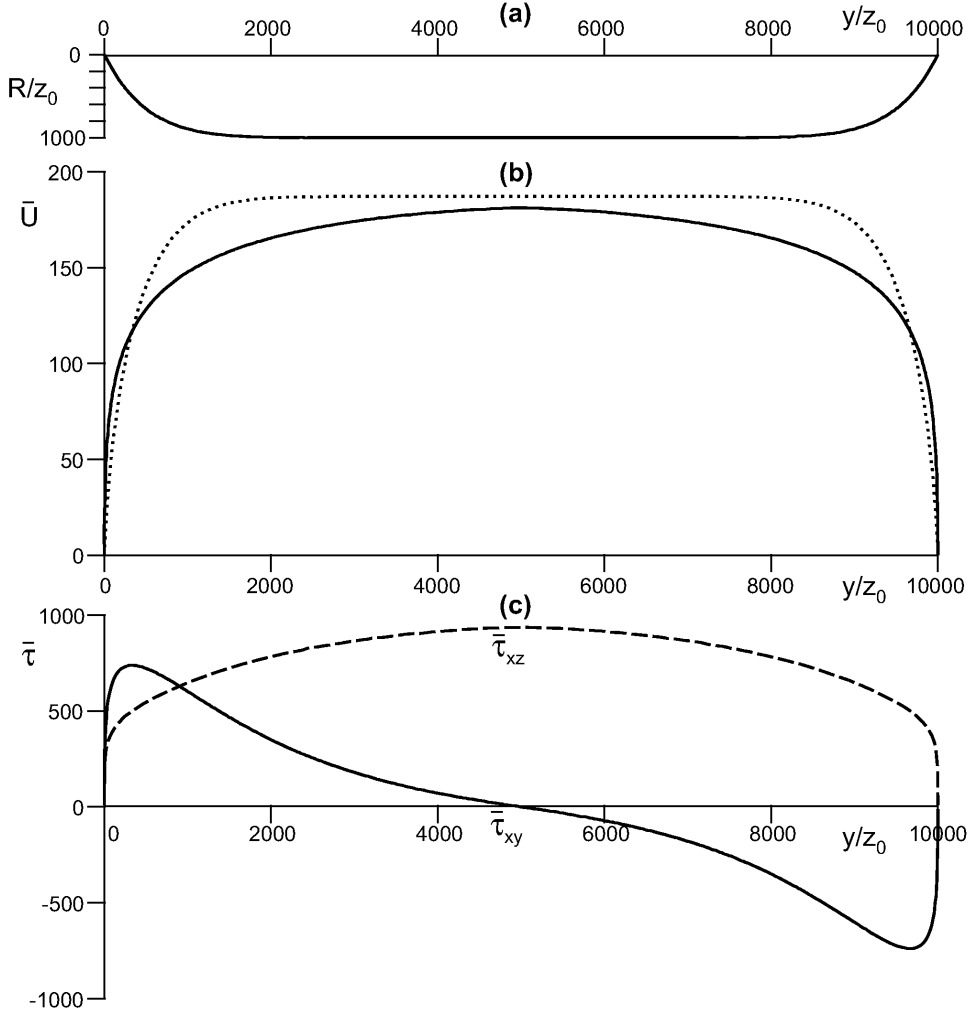}
\caption{Transverse profiles of (a) the river bed, (b) the mean flow velocity and (c) the two components $\tau_{xz}$ (long dashed line) and $\tau_{xy}$ (solid line) of the shear stress on the bed. Lengths are in units of the bed roughness $z_0=y_0$. The velocity is rescaled by $\sqrt{g z_0 \sin\theta/\kappa}$, and the stresses by $g z_0 \sin\theta/\kappa$. For this example, we choose $R(y) = D \left ( 4 \, y/W \right )^{10} \left (1 -  y/W \right )^{10}$ with channel deepest depth $D= 1000 z_0$ and channel width $W= 10 D$. In panel (b), the solid line corresponds to the velocity field computed taking into account the friction due to the banks. It is qualitatively different and quantitatively lower than that including the bed friction only (dashed line).}
\label{river} 
\end{figure}

To illustrate this method in the turbulent case, we study the transverse mean velocity profile in a river homogeneous along the streamwise direction. This is a standard hydrological geometry, usually treated within the shallow water approximation (\cite{J02}). We show here that the transverse momentum fluxes plays a significant role. In other words, the predicted mean flow is sensitive to the way friction on the bed and the banks is taken into account (\cite{FCA07}). As previously, we call $x$ the direction of the flow, $\tan\theta$ the slope of the river and $z$ the axis normal to the bed. $y$ designates the direction perpendicular to the flow. For this example, we take a simple polynomial form for the transverse depth profile $R(y)$, shown in figure \ref{river}(a). Such a flat and elongated shape is typically that of channels or river beds. Note that the above equations have been derived for a flow over a bottom varying along the \emph{streamwise direction} $x$ in order to allow for a straightforward comparison between the three cases (viscous, turbulent and granular flow rules). For the determination of the velocity profile in a river, the bed varies along  the \emph{direction transverse to the flow}, $y$. It thus requires a specific derivation of the integral method. As it strictly follows the same procedure, we shall mainly explicit the main steps and emphasise the differences. 

At first sight, such a channel geometry is close to the homogeneous turbulent boundary layer. This suggests that a Prandtl-like closure, although phenomenological, may be successful. However, four different distances come into play to construct a mixing length $\ell$: the distances to the bed and to the free surface as in the previous sub-sections (i.e. in the $z$ direction), but also the distances to the left and right banks of the river (in the $y$ direction). For the sake of simplicity, we decouple the $z$ and $y$ contributions by writing
\begin{eqnarray}
\tau_{xz} & = & \ell_z^2 |\dz u| \dz u
\quad \mbox{with} \quad
\ell_z = \kappa (z-Z+z_0) \sqrt{\frac{H-z}{R}} \, , \label{ellz} \\
\tau_{xy} & = & \ell_y^2 |\dy u| \dy u
\quad \mbox{with} \quad
\ell_y = \kappa' (\delta y+y_0)
\quad \mbox{and} \quad
\frac{1}{\delta y} = \frac{1}{y} + \frac{1}{W-y} \, . \label{elly}
\end{eqnarray}
In the above expression, $y_0$ is the roughness of the river banks --~we shall below simply take $y_0=z_0$ as well as $\kappa'=\kappa$. The choice of a so-called `harmonic mean' for the computation of $\delta y$ is simply a trick to get, in a smooth way, the smaller (and thus limiting) value among the distances to the banks $y$ and $W-y$. In this case, the water free surface is flat and is taken as the reference, so that $H=0$, i.e. $Z(y)=-R(y)$. Keeping again the relations (\ref{uTurbul}) and (\ref{pTurbul}) for the velocity and pressure profiles, one can then explicitly compute all stress components
\begin{eqnarray}
\tau_{xz} & = & C_z |U| U \, \frac{H-z}{R} \, , \\
\tau_{xy} & = & (\delta y + y_0)^2 \left [
\ln^2 \left ( 1+\frac{z-Z}{z_0} \right ) \left | \dy (\sqrt{C_z} U) \right | \dy (\sqrt{C_z} U)
\right . \nonumber \\
 & - &
\left .
2 \ln \left ( 1+\frac{z-Z}{z_0} \right ) \frac{\dy Z}{z-Z+z_0} \, \left | \dy (\sqrt{C_z} U) \right | \sqrt{C_z} U +
C_z \left ( \frac{\dy Z}{z-Z+z_0} \right )^2
\right ], \qquad
\end{eqnarray}
where we have introduced the notation $U = q/R$, so that the Saint-Venant equation reads:
\begin{eqnarray}
\dt q & = & gR \sin \theta - C_z |U| U +
\dy \left \{ (\delta y +y_0)^2 \left [ \zeta R \left | \dy (\sqrt{C_z} U) \right | \dy (\sqrt{C_z} U)
\right . \right . \nonumber \\
& + &
\left . \left .
\lambda^2 \left | \dy (\sqrt{C_z} U) \right | \dy R \, \sqrt{C_z} U +
\frac{R}{z_0(R+z_0)}  \left | \dy R \right | \dy R \, C_z |U| U  \right ] \right \} 
\nonumber \\
& + &
 \left (\frac{\delta y + y_0}{z_0}\right )^2 C_z \left | \dy R \right |^3  |U| U,
\end{eqnarray}
with $\zeta = (1+z_0/R)(\lambda^2-2\lambda)+2$.

This rather complicated equation can be greatly simplified under the assumptions that $\left | \dy R \right | \ll 1$ (weak transverse slope) and $z_0 \ll R$, which gives at leading order
\begin{equation}
\dt U = g \sin\theta - C_z \frac{U^2}{R} +
\dy \left \{ \kappa^2
\left [ y_0 + y \left ( 1 - \frac{y}{W} \right ) \right ]^2
\left | \dy U \right | \dy U
\right \}.
\label{U_river}
\end{equation}
It is important to note that the same expression is obtained if the logarithmic velocity profile is approximated by a plug flow, under the assumption $z_0 \ll R$ only --~it is a standard starting point in hydraulics. Whether or not the logarithmic profile is a better test function than the uniform one yet remains an open question that would require a specific study.

In the above expression, we can directly read that the last term is of diffusive form, with a turbulent eddy viscosity written in terms of $\ell_y$ and the mean velocity $U$ gradient. Without its contribution, i.e. without taking into account the friction of the banks, the steady solution of this equation is the usual hydrological relation (\cite{J02})
\begin{equation}
U_0=\sqrt{\frac{gR\sin\theta}{C_z}} \, .
\label{Uh}
\end{equation}
The result of the integration of equation (\ref{U_river}) is displayed in figure \ref{river}(b) with the solid line. The important point is that, although the typical bed profile curvature scales on the water depth, that of the velocity is more like the channel width. This is in agreement with experimental and field measures (\cite{FCA07}).  For comparison, the dashed line corresponds to the velocity computed with formula (\ref{Uh}). Its profile is, by contrast, as flat as $R(y)$. Finally, the stress profiles are shown in panel (c). As expected, the curvature of the curve $\tau_{xz}(y)$ is similar to that of $U(y)$. More interesting is the profile $\tau_{xy}(y)$: it goes very quickly (almost at the bank) to its maximum (absolute) value and then gently decreases down to zero at the center of the channel. This large value of $\tau_{xy}$ is of the order of that of $\tau_{xz}$ at the center of the channel where the flow velocity is the largest. This means that the banks are in fact as stressed as the center of the channel bed. This remark may be of importance as far as erosion processes are concerned.

\section{Dense granular flows}

The integral method can be applied to the case of a dense granular material flowing down an inclined plane, just like for a fluid. As for the above turbulent flow case, the starting equations are identical to (\ref{NS}) but the viscous term $\nu \Delta u$ is replaced by $\partial_j \tau_{xj} $. Of course, the constitutive relation, i.e. the relation between the stress tensor to the velocity field, has to be specified. We use here the recent advances in this field obtained by \cite{GDR04,dCEPRC05,JFP06}.

\subsection{Local rheology}
It has been shown that the rheology of dense granular flows is local in first approximation. This means that the stress and the shear rate at a given location in the flow are related through a univoque relation. From dimensional analysis the rheological law can be written as $\tau/p= \mu(I)$, where the friction coefficient $\mu$ is a function of the properly rescaled shear rate $I=|\dot{\gamma}| d/\sqrt{p}$ (\cite{GDR04}) and $d$ is the grain diameter. Following \cite{JFP06}, we split the stress tensor as $-p\,\delta_{ij} + \tau_{ij}$, where $p$ is the isotropic pressure, and write the friction law for a granular material as
\begin{equation}
\tau_{ij} = \frac{\mu(I)\,p}{|\dot{\gamma}|}\dot{\gamma}_{ij}
\label{defTauGranu}
\end{equation}
where $\dot{\gamma}_{ij}=\partial_iu_j+\partial_ju_i$ is the strain tensor, $|\dot{\gamma}| = \sqrt{\frac{1}{2}\dot{\gamma}_{ij}\dot{\gamma}_{ij}}$ is the second invariant of $\dot{\gamma}_{ij}$.

\subsection{Steady uniform solutions}
For the sake of simplicity, we shall restrict to a 2D geometry and look for the homogeneous and steady solutions. The momentum balance reads
\begin{equation}
0 = g\sin\theta + \partial_z \tau_{xz}.
\end{equation}
As the shear stress vanishes at the free surface of the dense granular flow, we get
\begin{equation}
\tau_{xz} = \mu(I) p = g(H_0-z)\sin\theta,
\end{equation}
where $H_0$ is the depth of the flowing layer. Similarly, the isotropic pressure increases linearly with depth: $p = g(H_0-z)\cos\theta$. The ratio of the two gives a constant friction coefficient: $\mu = \tan\theta$. As all other components of the strain tensor vanish, the shear rate $\dot{\gamma}_{xz}$ is then selected by the relationship:
\begin{equation}
I_0 = \frac{\dot{\gamma}_{xz}d}{\sqrt{p}} = \mu^{-1}(\tan\theta),
\end{equation}
where $\mu^{-1}$ denotes the inverse function of $\mu$. The integration along the $z$-axis of $\dot{\gamma}_{xz}$ leads to a Bagnold-like velocity profile
\begin{equation}
\frac{u}{\sqrt{gd}} = \frac{2}{3} I_0(\theta) \sqrt{\cos\theta} \, \frac{H_0^{3/2}-(H_0-z)^{3/2}}{d^{3/2}}.
\end{equation}
Note that this profile is in good agreement with recent two-dimensional numerical simulations of sheared granular layers (\cite{dCEPRC05}).

\subsection{Integral method}
At the sight of the above homogeneous steady solutions, we introduce the following test function for $u$
\begin{equation}
u = \frac{5}{3}\frac{q}{R} \left (\frac{R^{3/2}-(H-z)^{3/2}}{R^{3/2}} \right ),
\label{ugranu}
\end{equation}
and keep a hydrostatic pressure profile:
\begin{equation}
p = p_{\rm atm} + g (H-z)\cos\theta.
\label{pgranu}
\end{equation}
$R$, $H$ and $q$ denote the same quantities as in the previous sections. As for $\tau_{ij}$, we first need to compute $\dot{\gamma}_{ij}$. With the above expression of $u$, one could in principle compute $w$ from the continuity equation. Its expression is rather heavy, but under the assumption $u \gg w$ we get to the first order in $w/u$: $\dot{\gamma}_{xx} = - \dot{\gamma}_{zz} = 2 \dx u$ and $\dot{\gamma}_{xz} = \dz u$, so that the second invariant of the strain tensor reduces to $|\dot{\gamma}| = \dz u$. Finally, the stress tensor components read:
\begin{eqnarray}
\tau_{xx} = -\tau_{zz} & = & \mu(I) \, g \cos\theta \left [ \left ( \frac{4}{3q} \sqrt{R(H-z)} (R\dx q-q \dx R)
\right ) \right .\nonumber\\
& - & \left . 2 (H-z) \dx H + \frac{4}{qR} (H-z)^2 \left (\frac{5}{2}q\dx R -R\dx q\right )\right ],\\
\tau_{xz} & = & \mu(I) \, g(H-z)\cos\theta .
\end{eqnarray}

The effective friction coefficient $\mu(I)$ can be determined experimentally using homogeneous steady flows (\cite{GDR04, dCEPRC05,JFP06}). While several forms have been proposed for $\mu(I)$ (\cite{P99a, dCEPRC05,JFP06,A07}), the linear expansion around its static value $\tan\theta$,
\begin{equation}
\mu(I) = \tan\theta + m(I-I_0),
\end{equation}
is sufficient for our practical purpose. For the  glass beads used by \cite{P99a} ($d=500~\mu$m), the coefficient $m = d\mu/dI$ is around $0.5$. Within this linear approximation, the Saint-Venant equation gives:
\begin{equation}
\dt q + \dx \left (\alpha \frac{q^2}{R} \right ) = - gR\cos\theta [m(I-I_0) + \dx H] + \int_Z^H \dx \tau_{xx} \, dz
\end{equation}
where $\alpha = 5/4$ due to the Bagnold-like profile for $u$. Making the last term explicit, we can write the full set of equations for the dense granular case as
\begin{eqnarray}
Z + R & = & H \nonumber\\
\dt R + \dx q & = & 0 \nonumber \\
\dt q + \dx \left (\alpha \frac{q^2}{R} \right )
& = & - g R \cos \theta \left \{ m(I-I_0) + \dx H + \tan\theta \left [  R \left ( \frac{7}{9}\dxx R + \dxx Z \right ) \right . \right . \nonumber\\
& - & \left . \left . \frac{4}{9q} R^2 \dxx q \left (\frac{14}{9}(\dx R)^2+2\dx H \, \dx Z\right ) \right . \right . \nonumber\\
& + & \left . \left . \frac{4}{9q^2} R^2 (\dx q)^2 -\frac{4}{3q} R\dx q \, \dx R \right ] \right \}
\label{SVgranu}
\end{eqnarray}
%

\subsection{Example: the Kapitza instability}
Several experimental works report that surge waves deform the free-surface of granular flows down a rough inclined plane (\cite{S79,D01,FP03,MLAC06}). \cite{FP03} has shown that this instability is of the same nature as that observed in classical fluids (hence the name `Kapitza instability') but with specificities due to the granular rheology. We show here that our equations allow to recover the correct characteristics of this instability, lost with the long wavelength expansion derived by \cite{FP03}.

As the bed is now flat, the first equation of (\ref{SVgranu}) simply reduces to $R = H$. In the limit of small perturbations, we can look for a solution of the form:
\begin{equation}
H = H_0 \left ( 1 + H_1 e^{i(\bar{k} x/H_0-\bar{\omega} t \, q_0/H_0^2)} \right )
\quad \mbox{and} \quad
q = q_0 \left ( 1 + q_1 e^{i(\bar{k} x/H_0-\bar{\omega} t \, q_0/H_0^2))} \right ),
\end{equation}
where $\bar{k}=kH_0$ denotes the rescaled wavenumber and $\bar{\omega}=\omega H_0^2/q_0$ the dimensionless angular frequency. To the first order, the system of equation (\ref{SVgranu}) becomes:
\begin{eqnarray}
- i\bar{\omega} H_1 + i\bar{k} q_1 & = & 0, \nonumber \\
\Fr^2 \left[ \left (-\frac{1}{\alpha} i\bar{\omega} + 2 i\bar{k} \right ) q_1 - i\bar{k} H_1 \right ]
& = & \left ( \frac{7}{9}\tan\theta \bar{k}^2 + \frac{5}{2}m I_0 - i\bar{k} \right ) H_1 \nonumber \\
& - & \left (m I_0 + \frac{4}{9} \tan\theta \bar{k}^2 \right ) q_1,
\end{eqnarray}
where the Froude number of the flow is defined by:
\begin{equation}
\Fr^2 =  \alpha \, \frac{q_0^2}{g H_0^3 \cos\theta} \, .
\end{equation}
$\bar{k}$ and $\bar{\omega}$ are finally related by the implicit dispersion relation:
\begin{equation}
-\frac{1}{\alpha}\bar{\omega}^2 + 2\bar{\omega} \bar{k} + i\frac{m I_0}{\Fr^2} (\frac{5}{2}\bar{k} - \bar{\omega}) + (\frac{1}{\Fr^2}-1)\bar{k}^2 = i\frac{\tan\theta}{9\Fr^2}(4\bar{\omega}-7\bar{k})\bar{k}^2
\label{reldisp}
\end{equation}
By contrast to the dispersion relation of (\cite{FP03}), the above one now includes new terms -- those gathered on the right hand side. This instability being convective, a real $\omega$ is imposed and the instability develops spatially downflow. We then split the wave number into two terms $\bar{k} = \bar{\omega}/\bar{c} - i \bar{s}$, where $\bar{c}$ and $\bar{s}$ represent the (dimensionless) phase velocity and the spatial linear growth rate. Finally the functions $\bar{s}(\bar{\omega})$ and $\bar{c}(\bar{\omega})$ are computed from (\ref{reldisp}).

\begin{figure}
\centering
\includegraphics{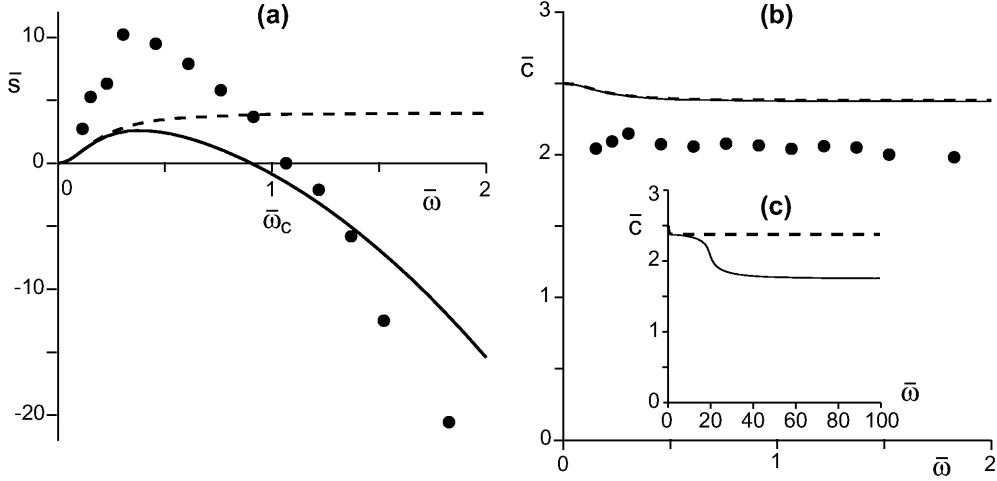}
\caption{\label{SigmaKGranu} Theoretical (lines) and experimental ($\bullet$) dispersion relation for $\theta = 29\deg$ and $\Fr = 1.14$. Spatial growth rate $\bar{s}$ (a) and phase velocity $\bar{c}$ (b) as functions of the pulsation $\bar{\omega}$. Inset (c) shows $\bar{c}$ for larger values of $\bar{\omega}$. Note the fall-off of the curve around $\bar{\omega} \sim 20$. The dashed lines represent theoretical curves of \cite{FP03} with $\alpha=\frac{5}{4}$ -- In their original paper they rather chose $\alpha=1$. Lengths  are in units of $H_0$ and times in units of $q_0/H_0^2$.}
\end{figure}

Equation (\ref{reldisp}) is of the third order and among the three solutions, two correspond to relaxation modes: the product $\bar{s} \bar{c}$ is negative for all $\bar{\omega}$ and all Froude numbers. The analysis of the third one shows that the flow is unstable when the Froude number exceeds the critical value $\Fr_c=1$ (independently of $\theta$). For $\Fr > 1$, small angular frequencies are unstable up to a cut-off value $\bar{\omega}_c$. This cut-off frequency vanishes at $\Fr=\Fr_c$, which means that it is a zero mode instability. The re-stabilization of the flow for frequencies larger than $\bar{\omega}_c$ is missed without the new terms in the dispersion relation (\ref{reldisp}).

As for this problem, experimental data (\cite{FP03}) as well as the results of a numerical integration of the full equations (\cite{F06}) are available, it is interesting to discuss some quantitative aspects of the integral method predictions. The results of our calculations as well as the data are displayed in figure \ref{SigmaKGranu}. In panel (a), one can see that, with $\Fr = 1.14$ corresponding to the experiments, the estimation of $\bar{\omega_c}$ is very good, however, the magnitude of the maximum growth rate $\bar{s}_{\rm max}$ is under-estimated by a factor of $3$. A similar quantitative difference is found for the value of the critical Froude number: its experimental determination is $\Fr_c \sim 0.6$, i.e. almost twice smaller than our prediction. For comparison, the computation of $\bar{s}_{\rm max}$ and $\Fr_c$ in \cite{F06} are respectively a factor of $2$ below and $10$\% above the corresponding experimental values. In panel (b) is displayed the velocity of the waves. The agreement between theory and experiment is fair. Interestingly, our calculation predicts a fall-off of the curve around $\bar{\omega} \sim 20$ and a cross-over to a second plateau, see panel (c). It is worth noting that this change of behaviour at large frequency is also predicted in the liquid case. This unexpected behaviour of the curve $\bar{c}(\bar{\omega})$ could not be checked with the calculations of \cite{F06} as, in accordance with the experimental data, the numerics was run up to $\bar{\omega}=2$ only.

\section{Summary and perspectives}

In this paper, we have shown that, for the description of the dynamical properties of flows, one can go beyond the lubrication or shallow water approximations following a well-defined procedure that can be adapted to any constitutive relation: the integral method. The equations derived within this framework are simpler to integrate than the full 3D equations (e.g. the Navier-Stokes equations) but possess the same physical ingredients. The technique consists in choosing `test functions' for the velocity profiles, whose functional form is inferred from the exact homogeneous and steady solutions of the full equations, and in performing a depth averaging procedure with these profiles.

We have successfully applied this method to situations as diverse as viscous, turbulent and granular flows down an incline. In these three cases, we have investigated a specific example, respectively, formation of anti-dunes, velocity profile in a river cross-section and development of the Kapitza instability. We have emphasised the benefits of the method compared to the usual approximations. In particular, we have shown that important physical mechanisms are qualitatively recovered: the presence of wavelength or frequency cut-offs in the dispersion relations of anti-dune or Kapitza instabilities, the significant role of the friction at the river banks on the velocity profile curvature.

A strong assumption in the derivation proposed concerns the exchanges of vertical momentum. In the three cases, we have indeed assumed that the pressure was dominated by gravity effects. It has for instance been shown by \cite{SH91} that the topography induces corrections to the hydrostatics pressure profile. It has also been shown by \cite{MBVLAP05} that, in the first stages of the vertical collapse of a granular column, gravity is not balanced by the pressure but by the vertical acceleration. The description of such a situation is beyond the scope of the equations derived here. However, the same strategy could be applied, using the depth-averaged vertical momentum equation to determine the pressure, whose profile would be prescribed by a test-function. By analogy to the choice made for the velocity,  a linear pressure profile is probably the best choice as one recovers the above equations when vertical inertia and vertical diffusion of momentum are negligible. A dedicated work is needed to test this idea further on. Our aim is now to use the integral method as a tool to investigate physical problems such as the formation of current ripples versus dunes in viscous or turbulent flumes, the river width selection or the description of fingering effects in granular avalanches.

\vspace*{0.5cm}
\noi
\rule[0.1cm]{3cm}{1pt}\\
We thank E. Cl\'ement and J.H. Snoeijer for a careful reading of the manuscript. We are grateful to Y. Forterre for providing his experimental data. This study was financially supported by an `ACI Jeunes Chercheurs' of the french ministry of research. 


\end{document}